\documentclass[aps,preprint,showkeys,showpacs]{revtex4}%
\usepackage{amsfonts}
\usepackage{amsmath}
\usepackage{amssymb}
\usepackage{graphicx}%
\begin{document}
\preprint{ }
\title[ ]{Upper bounds for fusion processes\\in collisions of weakly bound nuclei}
\author{L.F.~Canto, R.~Donangelo}
\affiliation{Instituto de F\'{\i}sica, Universidade Federal do Rio de Janeiro, C.P.68.528,
21941-972 Rio de Janeiro, Brazil}
\author{H.D.~Marta}
\affiliation{Instituto de F\'{\i}sica, Facultad de Ingenier\'{\i}a, C.C. 30, C.P. 11000
Montevideo, Uruguay}

\begin{abstract}
We obtain upper limits for the contributions of the incomplete fusion and
sequential complete fusion processes to the total fusion cross section.
Through those upper bounds we find that these processes are negligible in
reactions induced by projectiles such as $^{6}$He and $^{11}$Li, which break
up into neutrons and one fragment containing the full projectile charge.

\end{abstract}
\maketitle

The effects of channel coupling in fusion reactions induced by weakly bound
projectiles have attracted great interest over the last decade \cite{BCH93}.
Some theoretical studies predict strong influence of the breakup channel over
the complete fusion (CF) cross section \cite{Hu92,Ta93,DV94,Ha00,Di02,Di03}. 
When one tries to compare these predictions with experimental data
\cite{Da02,Si99,Si02,Ko98,Tr00,Al02,Re98}, one finds a serious problem.
Sorting out complete and incomplete fusion (IF) events in an experiment may 
be a very difficult task, specially when uncharged fragments are produced in 
the breakup of the weakly bound collision partner. For this reason, most
experiments measure the total fusion cross section, $\sigma_{TF}=\sigma
_{CF}+\sigma_{IF}.$\ These results could not, in principle, be used to check
theoretical predictions for $\sigma_{CF}.$ However, the situation would be
different when $\sigma_{IF}<<\sigma_{CF}.$ In this case one can approximate
$\sigma_{TF}\simeq\sigma_{CF}$ and the measured cross section could be
directly compared with theoretical predictions for $\sigma_{CF}.$ In the
present work, we present a method to find upper limits for $\sigma_{IF}$ in
collisions induced by weakly bound projectiles. With this method, we show that
the incomplete fusion cross sections may be neglected when the projectile
breakup produces uncharged fragments.

The appropriate theoretical tool to handle this problem is the coupled-channels
method. However, its implementation becomes very complicated for the breakup
channel, since it involves an infinite number of states in the continuum. For
practical purposes, it is necessary to approximate the continuum by a finite
set of states as in the Continuum Discretized Coupled-Channels method (CDCC)
\cite{NT98}. This procedure has been extended to the case of fusion reactions
in refs.~{\cite{Ha00,Di02,Di03}}. Recently, a semiclassical alternative based
on the classical trajectory approximation of Alder and Winther (AW) \cite{AW}
has been proposed \cite{AW1}. This approximation was used to calculate breakup
cross sections and the results were compared with those of the CDCC method.
The agreement between these calculations was very good. Since this
semiclassical version of the CDCC method is much simpler, it may be a very
useful tool to calculate cross sections for other channels in reactions with
weakly bound nuclei. Although the AW method has been extensively used for
several nuclear reaction processes, only very recently it was applied to the
estimate of the fusion cross section \cite{Hu04}. For this application it was
considered a simplified two-channel problem for which the fusion cross section
obtained with the AW method was compared with results of a full
coupled-channels calculation. In spite of the large simplification in the
calculation the agreement between these two calculations was again very good.
Although such calculations may not be reliable for quantitative predictions,
they lead to a very useful qualitative conclusion. At above-barrier energies,
the fusion probability through channel-$\alpha$ at the partial-wave $l$ can be
written as a product of two factors. The first is the population of channel
$\alpha,$\ $\bar{P}_{l}^{\left(  \alpha\right)  },$ at the point of closest
approach. The second is the tunneling probability, $~T_{l}^{(\alpha)},$
through the effective ($l$-dependent) potential barrier. When dealing with
the breakup channel, one should have in mind that different tunneling factors
should be used for incomplete fusion of each breakup fragment. This point is
not considered when one treats the breakup channel as a bound state. A
quantitative semiclassical calculation of the fusion cross sections in
reactions with weakly bound projectiles requires the inclusion of the
continuum states associated with the breakup channel, as in ref. \cite{AW1}.
However, some simple upper bounds can be easily obtained.

As this work is devoted to reactions induced by weakly bound projectiles, the
variables employed to describe the collision are the projectile-target
separation vector, $\mathbf{r}$, and the relevant intrinsic degrees of freedom
of the projectile, $\xi$. For simplicity, we neglect the internal structure of
the target. The Hamiltonian is then given by%
\begin{equation}
h=h_{0}(\xi)+V(\mathbf{r},\xi), \label{h}%
\end{equation}
where $h_{0}(\xi)$ is the intrinsic Hamiltonian and $V(\mathbf{r},\xi)$
represents the projectile-target interaction. The eigenvectors of $h_{0}(\xi
)$\ are given by the equation%
\begin{equation}
h_{0}~\left\vert \phi_{\alpha}\right\rangle =\varepsilon_{\alpha}~\left\vert
\phi_{\alpha}\right\rangle . \label{av}%
\end{equation}
The AW method \cite{AW} is implemented in two-steps. First, one employs
classical mechanics for the time evolution of the variable $\mathbf{r}$. The
ensuing trajectory depends on the collision energy, $E,$ and the angular
momentum, $l$. In its original version, an energy symmetrized Rutherford
trajectory $\mathbf{r}_{l}(t)$ was used. In our case, the trajectory is the
solution of the classical equations of motion with the potential
$V(\mathbf{r)=}$\ $\left\langle \phi_{0}\right\vert V(\mathbf{r}%
,\xi)\left\vert \phi_{0}\right\rangle ,$ where $\left\vert \phi_{0}%
\right\rangle $\ is the ground state of the projectile. In this way, the
coupling interaction becomes a time-dependent interaction in the $\xi$-space,
$V_{l}(\xi,t)\equiv V(\mathbf{r}_{l}(t),\xi).$ 
The second step consists of treating the dynamics in the intrinsic space as 
a time-dependent quantum mechanics problem. Expanding the wave function in 
the basis of intrinsic eigenstates,%
\begin{equation}
\psi(\xi,t)=\sum_{\alpha}a_{\alpha}(l,t)~\phi_{\alpha}(\xi)~e^{-i\varepsilon
_{\alpha}t/\hbar}, \label{exp}%
\end{equation}
and inserting this expansion in the Schr\"{o}dinger equation for $\psi
(\xi,t),$ one obtains the AW's equations%
\begin{equation}
i\hbar~\dot{a}_{\alpha}(l,t)=\sum_{\beta}~\left\langle \phi_{\alpha
}\right\vert V_{l}(\xi,t)\left\vert \phi_{\beta}\right\rangle
~e^{i\left(  \varepsilon_{\alpha}-\varepsilon_{\beta}\right)  t/\hbar
}~a_{\beta}(l,t). \label{AW}%
\end{equation}
These equations are solved with the initial conditions $a_{\alpha
}(l,t\rightarrow-\infty)=\delta_{\alpha0},$ which means that before the
collision ($t\rightarrow-\infty$) the projectile was in its ground state. The
final population of channel $\alpha$ in a collision with angular momentum $l$
is $P_{l}^{(\alpha)}=\left\vert a_{\alpha}(l,t\rightarrow+\infty)\right\vert
^{2}$ and the angle-integrated cross section is
\begin{equation}
\sigma_{\alpha}=\frac{\pi}{k^{2}}~\sum_{l}\left(  2l+1\right)  ~P_{l}%
^{(\alpha)}. \label{PW1}%
\end{equation}

To extend this method to fusion reactions, we start with the quantum
mechanical calculation of the fusion cross section in a coupled channel
problem. For simplicity, we assume that all channels are bound and have spin
zero. The fusion cross section is a sum of contributions from each channel.
Carrying out partial-wave expansions we get%

\begin{equation}
\sigma_{TF}=\sum_{\alpha}~\left[  \frac{\pi}{k^{2}}\sum_{l}\left(
2l+1\right)  ~P_{l}^{F}(\alpha)\right]  , \label{sigTF}%
\end{equation}
with%
\begin{equation}
P_{l}^{F}(\alpha)=\frac{4k}{E}~\int dr~\left\vert u_{\alpha l}(k_{\alpha
},r)\right\vert ^{2}~W_{\alpha}^{F}(r). \label{PQM}%
\end{equation}
Above, $u_{\alpha l}(k_{\alpha},r)$ represents the radial wave function for
the $l^{th}$-partial-wave in channel $\alpha$ and $W_{\alpha}^{F}$ is the
absolute value of the imaginary part of the optical potential associated to fusion.

\bigskip

To use the AW method to evaluate the complete fusion cross section, we make
the approximation%
\begin{equation}
P_{l}^{F}(\alpha)\simeq\bar{P}_{l}^{\left(  \alpha\right)  }~T_{l}^{(\alpha
)}(E_{\alpha}). \label{PLAW}%
\end{equation}
Above, $T_{l}^{(\alpha)}(E_{\alpha})$\ is the\ probability\ that a particle
with reduced mass $\mu_{\alpha}=m_{0}A_{P}A_{T}/\left(  A_{P}+A_{T}\right)
$\ and energy $E_{\alpha}=E-\varepsilon_{\alpha}$\ tunnels through the
potential barrier in channel $\alpha$, and $\bar{P}_{l}^{\left(
\alpha\right)  }$ is the probability that the system is in channel-$\alpha\ $
at the point of closest approach on the classical trajectory.

We now proceed to study the complete and incomplete fusion cross sections in
reactions induced by weakly bound projectiles. For simplicity, we assume that
the GS is the only bound state of the projectile (as is the case of $^{11}$Li
projectiles) and that the breakup process produces only two projectile
fragments, $F_{1}$ and $F_{2}$. In this way, the labels $\alpha=0$ and 
$\alpha\neq0$ correspond respectively to the GS and the breakup states 
represented by two unbound fragments. Neglecting any sequential contribution, 
the complete fusion can only arise from the elastic channel. In this way, the 
cross section $\sigma_{CF}$\ can be obtained from eq.(\ref{sigTF}), dropping 
the sum over channels and using in the single term
\begin{equation}
\bar{P}_{l}^{\left(  0\right)  }\equiv P_{l}^{Surv}=\left\vert a_{0}%
(l,t_{ca})\right\vert ^{2}. \label{Psurv}%
\end{equation}
This probability is usually called survival (to breakup) probability. We get%

\begin{equation}
\sigma_{CF}=\frac{\pi}{k^{2}}\sum_{l}\left(  2l+1\right)  ~P_{l}^{Surv}%
~T_{l}^{(0)}(E). \label{sigCF}%
\end{equation}

The accuracy of the semiclassical fusion cross section has recently been
checked in a preliminary two-channel calculation in the scattering of $^{6}$He
projectiles on a $^{238}$U target, at near barrier energies \cite{Hu04}. The
weakly bound $^{6}$He nucleus dissociates into $^{4}$He and two neutrons, with
threshold energy $B=0.975$ MeV. The elastic channel is strongly coupled to the
breakup channel and the influence of this coupling on the fusion cross section
is very important. In this model, the breakup channel is represented by a
single effective state~\cite{Ca03}. For simplicity, the effective channel is
treated as a bound state but it is assumed to contribute only to incomplete
fusion. The complete fusion cross section is therefore given by
eq.(\ref{sigCF}) and the incomplete fusion cross section by considering only
the $\alpha=1$ term in eqs.(\ref{sigTF}) and (\ref{PLAW}). In \cite{Hu04} the
threshold energy was neglected and the same potential barrier was used for
both channels. That work showed that above the Coulomb barrier the
semiclassical cross sections (both $\sigma_{CF}$ and $\sigma_{TF})$\ are in
very good agreement with those calculated with the coupled-channels method.
Further evidences of this fact will be presented in a forthcoming paper
\cite{Longpaper}.%
%TCIMACRO{\FRAME{ftbpFU}{4.3543in}{4.7504in}{0pt}{\Qcb{Total fusion cross
%section of ref.\cite{Hu04} (solid squares) compared with that of the present
%work (stars). The present calculation uses the same potential, channel
%coupling and simplifying assumptions of \cite{Hu04}. The basic difference is
%that here the contribution from incomplete fusion uses the tunneling of the
%$^{4}$He fragment, rather than the full $^{6}$He projectile. For comparison,
%the complete fusion cross section of \cite{Hu04} is also shown (open
%squares).}}{}{figure1.eps}{\special{ language "Scientific Word";
%type "GRAPHIC";  maintain-aspect-ratio TRUE;  display "USEDEF";
%valid_file "F";  width 4.3543in;  height 4.7504in;  depth 0pt;
%original-width 8.0756in;  original-height 8.2339in;  cropleft "0";
%croptop "0.8388";  cropright "1";  cropbottom "0.0805";
%filename 'figure1.eps';file-properties "XNPEU";}}}%
%BeginExpansion
\begin{figure}
[ptb]
\begin{center}
\includegraphics[
trim=0.000000in 0.662829in 0.000000in 1.327305in,
height=4.7504in,
width=4.3543in
]%
{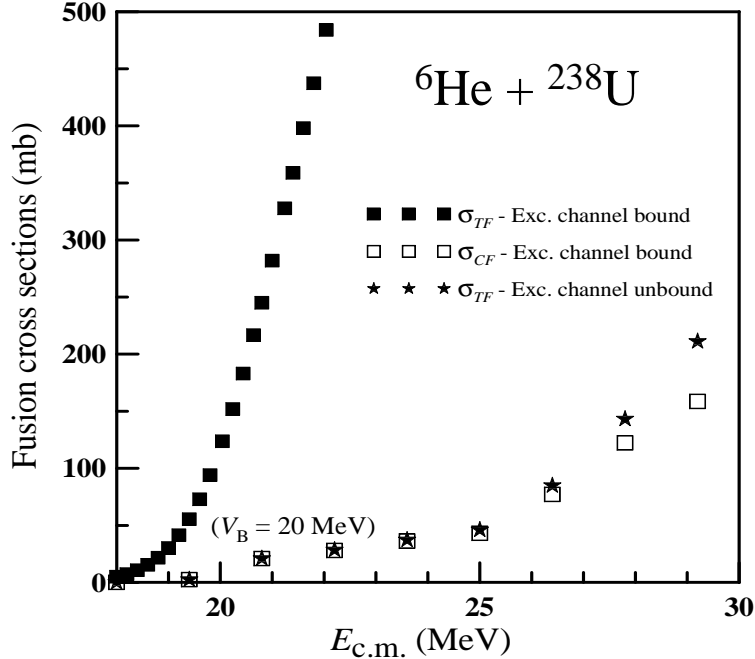}%
\caption{Total fusion cross section of ref.~\cite{Hu04} (solid squares)
compared with that of the present work (stars). The present calculation uses
the same potential, channel coupling and simplifying assumptions of
\cite{Hu04}. The basic difference is that here the contribution from
incomplete fusion uses the tunneling of the $^{4}$He fragment, rather than the
full $^{6}$He projectile. For comparison, the complete fusion cross section of
\cite{Hu04} is also shown (open squares).}%
\end{center}
\end{figure}
%EndExpansion

These calculations are rather schematic, since the continuum is represented by
a single bound effective channel. In this way $T_{l}^{(1)}(E_{1})$ is the
tunneling probability of the projectile through the projectile-target
potential barrier. However, incomplete fusion does not correspond to this
process. It corresponds to the tunneling of a projectile's fragment through
its barrier with respect to the target. In the particular collision studied in
\cite{Hu04}, that is $^{6}$He - $^{238}$U, incomplete fusion corresponds to
the fusion of $^{4}$He with $^{238}$U. The $^{4}$He fragment carries about 2/3
of the incident energy while the $^{4}$He-$^{238}$U\ potential barrier is
slightly higher then that for the entrance channel. Thus it is clear
that the incomplete fusion cross section is overestimated in our previous work
\cite{Hu04}. To illustrate this situation, in figure 1 we show the total
fusion cross section (solid squares) of \cite{Hu04} where in $\sigma_{TF}$ the
incomplete fusion contribution was obtained from eq.(\ref{sigTF}) with
$T_{l}^{(1)}(E_{1})$ representing the projectile-target tunneling probability.
We then re-calculate $\sigma_{TF}$ modifying the contribution from incomplete
fusion. We use the same $\bar{P}_{l}^{(1)}$ but replace the tunneling factor
by that for the $^{4}$He fragment. That is, we use the $^{4}$He - $^{238}%
$U\ potential barrier and the energy and angular momentum corresponding to the
shares of $^{4}$He in the $^{6}$He projectile. For simplicity, we neglect the
relative motion of the fragments of $^{6}$He. The resulting $\sigma_{TF}$ is
shown in figure 1 as stars. It is clear that a proper treatment of the
tunneling factor leads to a substantial reduction of $\sigma_{TF}.$ The new
cross section now is close to the complete fusion cross section $\sigma_{CF}$
also obtained in \cite{Hu04} (open squares). This indicates that the
incomplete fusion cross section $\sigma_{IF}$ is very small.

As we mentioned before, the above results cannot be considered as a realistic
prediction of the total fusion cross section, since the model does not use a
realistic description of the continuum states corresponding to the breakup
channel. Nevertheless we will show that such simple calculations are capable
of yielding relevant information on the fusion process: more precisely, upper
bounds for the incomplete fusion and the sequential complete fusion cross
sections, $\sigma_{IF}$ and $\sigma_{SCF}$, respectively, can be obtained from
eq.(\ref{sigTF}) setting $\bar{P}_{l}^{(1)}=1$ and evaluating the tunneling
probability in a proper way, as discussed below.

To illustrate the application of this procedure, we show two examples. We
employ the Aky\"{u}z-Winther parametrization for the interaction potentials
for all the systems considered. Furthermore, the ingoing wave boundary
condition is used in all these calculations. Note that in the schematic model
of figure 1 we neglected the breakup threshold energy. However, in the
following estimates of upper limits for the fusion cross sections, we do take
it into account.%
%TCIMACRO{\FRAME{ftbpFU}{4.3543in}{4.4469in}{0pt}{\Qcb{Upper bounds of the
%contributions to the incomplete fusion cross section for the $^{7}$Li +
%$^{209}$Bi system, employing the Aky\"{u}z-Winther parametrization for the
%interaction potentials. }}{}{figure2.eps}%
%width 4.3543in;  height 4.4469in;  depth 0pt;  original-width 7.8646in;
%original-height 11.3213in;  cropleft "0";  croptop "0.8387";  cropright "1";
%cropbottom "0.1289";  filename 'figure2.eps';file-properties "XNPEU";}}}%
%BeginExpansion
\begin{figure}
[ptb]
\begin{center}
\includegraphics[
trim=0.000000in 1.459316in 0.000000in 1.826126in,
height=4.4469in,
width=4.3543in
]%
{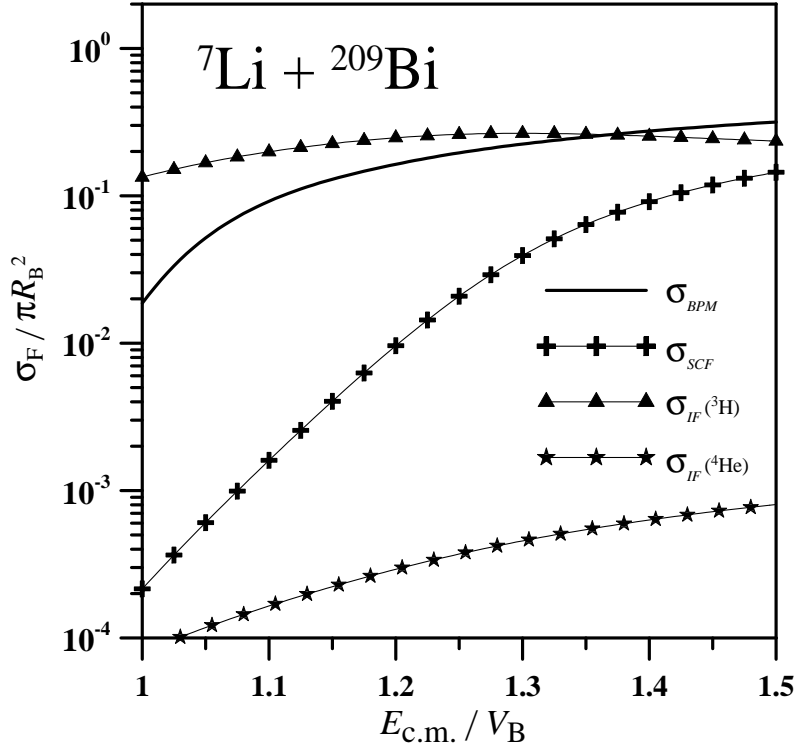}%
\caption{Upper bounds of the contributions to the incomplete fusion cross
section for the $^{7}$Li + $^{209}$Bi system, employing the Aky\"{u}z-Winther
parametrization for the interaction potentials. }%
\end{center}
\end{figure}
%EndExpansion

In the first case, shown in figure 2, we consider different fusion processes
that appear for the case of a $^{7}$Li projectile incident on a $^{209}$Bi
target, at energies just above the Coulomb barrier. Only energies above the
barrier are shown, as this is the region of applicability of the present
version of the method employed here \cite{Hu04}. The cross section for the
incomplete fusion induced by the $^{3}$H fragment is much larger than that for
$^{4}$He, which is negligible. This situation should be expected because of
the lower Coulomb barrier energy for $^{3}$H. Also shown is the single barrier
penetration model cross section, $\sigma_{BPM}$, for $^{7}$Li. 
We note that the upper bound
for the incomplete fusion cross section induced by the $^{3}$H fragment is
large, exceeding $\sigma_{BPM}$ in the low energy region. The experimental
findings for this system \cite{Da02} yield a value of the incomplete fusion
cross section of about 30\% of the total fusion cross section. Thus, although
our upper bound is compatible with the data, not much is learnt in this case.
Also shown in this figure is the upper bound for the cross section for
sequential complete fusion, $\sigma_{SCF}$. Although negligible at low
energies, it becomes appreciable for $E_{c.m.}/V_{B}\approx1.5$. We should
remark that to neglect the relative motion between the fragments tends to
overestimate the sequential complete fusion cross section, and to decrease our
estimate of the incomplete fusion cross sections. A quantitative investigation
of these effects is under way \cite{Longpaper}.

In the case of $^{6}$He incident on $^{238}$U shown in figure 3, only the
contribution from $^{4}$He to the incomplete fusion cross section must be
included, as the capture of one or both of the neutrons produced in the
breakup of $^{6}$He cannot be experimentally distinguished from the transfer 
process. In this
case the upper bound for both the incomplete fusion cross section, and the
sequential complete fusion cross sections are much smaller than the BPM
estimate for the complete fusion cross section. This shows that, although it
is difficult in this case to distinguish between the complete and total fusion
cross sections, their difference is expected to be small, as the value of the
incomplete fusion contributions to the total fusion cross section is not
important.%
%TCIMACRO{\FRAME{ftbpFU}{4.3543in}{3.8242in}{0pt}{\Qcb{Same as figure~1 for the
%$^{6}$He + $^{209}$Bi system. Note that only the $^{4}$He contribution to the
%incomplete fusion has been shown. See text for details and further
%discussion.}}{}{figure3.eps}{\special{ language "Scientific Word";
%type "GRAPHIC";  maintain-aspect-ratio TRUE;  display "USEDEF";
%valid_file "F";  width 4.3543in;  height 3.8242in;  depth 0pt;
%original-width 7.8646in;  original-height 11.3213in;  cropleft "0";
%croptop "0.7707";  cropright "1";  cropbottom "0.1611";
%filename 'figure3.eps';file-properties "XNPEU";}}}%
%BeginExpansion
\begin{figure}
[ptb]
\begin{center}
\includegraphics[
trim=0.000000in 1.823861in 0.000000in 2.595974in,
height=3.8242in,
width=4.3543in
]%
{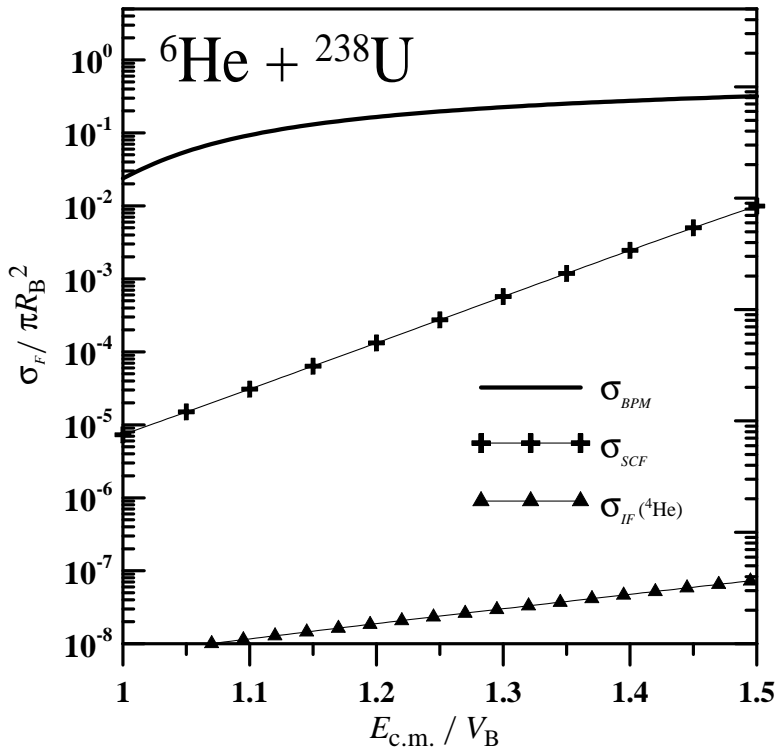}%
\caption{Same as figure~2 for the $^{6}$He + $^{209}$Bi system. Note that only
the $^{4}$He contribution to the incomplete fusion has been shown. See text
for details and further discussion.}%
\end{center}
\end{figure}
%EndExpansion

\bigskip

\bigskip In summary, we have illustrated how the application of the upper
bounds to the incomplete fusion cross sections may be applied to the estimate
of their contribution to the total fusion cross section. In cases where the
unstable nucleus breaks into charged fragments, these upper bounds are
consistent with the values measured. When one of the fragment posseses all of
the charge of the unstable nucleus, we have shown that the complete fusion
cross section, which is easy to evaluate theoretically, is a good estimate of
the measured total fusion cross section. The calculations presented here are
limited to energies above the Coulomb barrier. An extended version of the
method exploring the classically forbidden region and including the relative
motion between the fragments is presently being developed~\cite{Longpaper}.

\bigskip

We acknowledge useful discussions with P.R. Silveira Gomes. This work was
supported in part by MCT/FINEP/CNPq(PRONEX) under contract no. 41.96.0886.00,
PROSUL and FAPERJ (Brazil), and from PEDECIBA and CSIC (Uruguay).

\end{document}